\documentclass[aps,floatfix,nofootinbib,preprintnumbers]{revtex4}
\usepackage{amsmath}
\usepackage{amssymb}
\usepackage{epsfig}
\usepackage{graphicx}
\usepackage{rotating}
\usepackage{pstricks}
\usepackage{pst-coil}
\DeclareMathAlphabet{\mathsc}{OT1}{cmr}{m}{sc}
\newcommand {\ignore}[1]{}

\usepackage{color} %This is essential for placing graphics onto
\usepackage{multirow} % To use multirow command in tables
\usepackage[T1]{fontenc} 
\usepackage{slashed}

\def\10{$SO(10)$}
\def\21{SU(2) $\otimes$ U(1) }

\def\422{$SU(4) \otimes SU(2) \otimes SU(2)$}
\def\321{SU(3) $\otimes$ SU(2) $\otimes$ U(1)}
\def\gsim{\raise0.3ex\hbox{$\;>$\kern-0.75em\raise-1.1ex\hbox{$\sim\;$}}}
\def\lsim{\raise0.3ex\hbox{$\;<$\kern-0.75em\raise-1.1ex\hbox{$\sim\;$}}}

\def\lsim{\raise0.3ex\hbox{$\;<$\kern-0.75em\raise-1.1ex\hbox{$\sim\;$}}}
\def\gsim{\raise0.3ex\hbox{$\;>$\kern-0.75em\raise-1.1ex\hbox{$\sim\;$}}}
\def\vev#1{\left\langle #1\right\rangle}
\def \znbb {0\nu\beta\beta}

\newcommand{\ba}{\begin{array}}
\newcommand{\ea}{\end{array}}
\relax

\def\321{$SU(3)\times SU(2)\times U(1)$}

\newcommand{\AddWurz}{ 
 Institut f{\"u}r Theoretische Physik und
  Astrophysik, \\ Universit{\"a}t W{\"u}rzburg, 97074 W{\"u}rzburg,
  Germany}

\newcommand{\AddFrascati}{
INFN, Laboratori Nazionali di Frascati,
Via Enrico Fermi 40,    I-00044 Frascati, Italy}

\newcommand{\AddrOrsay}{%
Laboratoire de Physique Th\'eorique, CNRS -- UMR 8627, \\Universit\'e de Paris-Sud 11, F-91405 Orsay Cedex, France
}

\def\gsim{\raise0.3ex\hbox{$\;>$\kern-0.75em\raise-1.1ex\hbox{$\sim\;$}}}
\def\lsim{\raise0.3ex\hbox{$\;<$\kern-0.75em\raise-1.1ex\hbox{$\sim\;$}}}

\begin{document}

%\preprint{LPT-12-121}

\renewcommand{\Huge}{\Large}
\renewcommand{\LARGE}{\Large}
\renewcommand{\Large}{\large}
\def \znbb {$0\nu\beta\beta$ }
\def \nbb {$\beta\beta_{0\nu}$ }
\title{Flavor origin of R-parity}  \date{\today}

\author{S. Morisi} \email{stefano.morisi@gmail.com}
%\affiliation{\AddrAHEP}
\affiliation{\AddWurz}
\author{E. Peinado} \email{epeinado@lnf.infn.it}
%\affiliation{\AddrAHEP}
\affiliation{\AddFrascati}
\author{A. Vicente}\email{avelino.vicente@th.u-psud.fr}
\affiliation{\AddrOrsay}

\begin{abstract}
Proton stability is guaranteed in the MSSM by assuming a discrete
symmetry, R-parity. 
However, there are additional R-parity conserving higher
dimensional operators which violate lepton and baryon numbers and
induce fast proton decay. Here we study the possibility that all
renormalizable, as well as the most dangerous non-renormalizable,
R-parity violating operators are forbidden by a flavor symmetry,
providing a common origin for fermion mixing and proton and dark
matter stability.  We propose a specific model based on the
$\Delta(27)$ discrete symmetry.

\end{abstract}

%\pacs{
%11.30.Hv       % Flavor symmetries
%14.60.-z       % Leptons
%14.60.Pq       % Neutrino mass and mixing
%14.80.Cp       % Non-standard-model Higgs bosons
%23.40.Bw       % Weak-interaction and lepton (including neutrino) aspects of decay
%}

\maketitle

\section{Introduction}
\label{sec:intro}

In July 2012, the ATLAS \cite{ATLAStalk} and CMS \cite{CMStalk}
collaborations announced the discovery of a new boson, with a mass in
the $126$ GeV ballpark. Although detailed studies of its properties
need to be done in order to confirm its identity, one is tempted to
interpret this new particle in terms of the long-awaited Higgs
boson. In any case, this constitutes a decisive breakthrough in high
energy physics that, once correctly understood, will surely shed light
on the dynamics of the electroweak scale.
%\\\indent
The discovery of the Higgs boson, and the measurement of its mass,
reminds us a long-standing theoretical problem in particle physics:
the famous hierarchy problem~\cite{Gildener:1976ai}. This can be
expressed as the high sensitivity that fundamental scalars have to
physics at high energies. Unless one accepts a very precise
fine-tuning of the parameters of the theory, the Higgs mass is
naturally pushed to those high energies, distabilizing the electroweak
scale.
%\\\indent
Supersymmetry (SUSY) is one of the most popular solutions to the
hierarchy problem. If the SUSY breaking scale is low (not far from the
TeV scale), the electroweak scale is kept under control by virtue of
the cancellations between bosonic and fermionic contributions to the
Higgs mass.
\\\indent
When constructing a supersymmetric model one finds new gauge and SUSY
invariant renormalizable interactions, not present in the standard
model, that lead to lepton (L) and baryon (B) number violation. With
the particle content of the Minimal Supersymmetric Standard Model
(MSSM), these are
\begin{equation}
[L H_u]_F,\quad  
[L L e]_F,\quad
[L Q d]_F,\quad
[u d d]_F,\label{rpvt}
\end{equation}
where $F$ stands for F-terms.  If they were simultaneously present in
the lagrangian, the proton would have a fast decay rate unless very
small coefficients are introduced. For that reason, one usually
introduces a discrete symmetry called R-parity, $R_p =
(-1)^{3(B-L)+2s}$ (where $s$ is the spin of the particle), that
forbids the L and B violating terms shown above~\cite{Farrar:1978xj}.
The conservation of R-parity has very relevant phenomenological
implications. This discrete symmetry stabilizes the lightest
supersymmetric particle (LSP). As a consequence of that,
supersymmetric events at colliders contain large amounts of missing
energy in the final state. Furthermore, if neutral, the LSP would be
the perfect example of a Weakly Interacting Massive Particle (WIMP)
and a good dark matter candidate.
R-parity has, however, some drawbacks. First of all, R-parity is
introduced by hand in the MSSM, without a theoretical explanation for
its origin. And second, there are additional non-renormalizable
interactions which, even though they break lepton and/or baryon
numbers, are perfectly allowed by R-parity. Even if these operators
are generated at the Planck scale, they would lead to unnacceptably
fast proton decay unless their coefficients are
tiny~\cite{Weinberg:1981wj,Nath:2006ut}.
\\\indent
Many theoretical ideas have been proposed in order to explain the
origin of R-parity. Most of them consider R-parity as a remnant after
the breaking of a larger symmetry group, see for example
\cite{Mohapatra:1986su}. However, in most cases there is
no explanation for the suppression of the higher-dimensional
operators.
On the other hand, the origin of fermion masses hierarchies and
mixings, the so-called flavor problem, is another long-standing
mystery in particle physics.  One interesting possibility to address
the flavor problem is the introduction of an horizontal symmetry
between the three generations of fermions. The flavor symmetry, which
can be either continuous or discrete, imposes some structures on the
Yukawa couplings.  

In this paper we propose that R-parity may be a consequence of the
flavor symmetry. In this case, quark and lepton mixings, as well as
the stability of proton and dark matter, can be explained in a common
framework. We introduce a flavor model where all renormalizable as
well as dimension five L or B violating operators are forbidden by the
non-abelian flavor symmetry $\Delta(27)$. Charged femion masses and
mixings can be fitted to their observed values. Since the Weinberg
operator is also forbidden by the flavor symmetry, one is forced to go
beyond minimal models if neutrino masses and mixings are to be
explained. This turns out be a non-trivial task due to (1) the
restrictions imposed by $\Delta(27)$, and (2) our main motivation of
not inducing R-parity violating terms. We find that this can be
achieved by extending the lepton sector without inducing dangerous
dimension five operators.
%In particular we have
%  to include one extra scalar MSSM-singlet and two sets of fermion
%  electroweak triplets.
Neutrino mass arises from a variation of the inverse seesaw
mechanism~\cite{Mohapatra:1986bd}, called inverse type-III seesaw
mechanism~\cite{Ma:2009kh}, that provides a correct description of
neutrino mixings.

The rest of the paper is organized as follows: we present the model
and main idea in section \ref{sec:model}: how the introduction of a
$\Delta(27)$ flavor symmetry can automatically lead to R-parity
conservation and proton stability. Afterwards, we discuss the flavor
structure of the resulting quark and lepton sectors in sections
\ref{sec:quarksector} and \ref{sec:leptonsector}, respectively. In the
latter case we extend the original model in order to account for
neutrino masses and show how the resulting neutrino mixing pattern can
accommodate the data coming from oscillation experiments. Finally, we
conclude with a short summary and a brief discussion in section
\ref{sec:summary}.

\section{The model: R-parity and proton stability from $\Delta(27)$}
\label{sec:model}

Let us consider the MSSM extended by a $\Delta(27)$ flavor
symmetry, see for instance \cite{Luhn:2007uq}. This discrete group is a subgroup of $SU(3)$ (for a
classification see~\cite{Ishimori:2010au}) that belongs to the series
$\Delta(3 n^2)$. It has $11$ irreducible representations, namely two
triplets ${\bf 3}$, ${\bf 3}^*$ and 9 singlets ${\bf
  1_i}$. The product rules for the triplet representations are ${\bf 3}\times {\bf
  3}^*=\sum_{i=1}^9 {\bf 1_i}$ and ${\bf 3}\times {\bf
  3}={\bf 3}^*+{\bf 3}^*+{\bf 3}^*$. From
these rules, it is clear that the product ${\bf 3}\times {\bf 3}\times
{\bf 3}$ is invariant under $\Delta(27)$ whereas ${\bf 3}\times {\bf
  3}\times {\bf 3}^*$ is not.
We assign $\Delta(27)$ representations to the MSSM particle content as
shown in Table \ref{model}. It is straightforward to check that this
assignment forbids all the aforementioned R-parity violating
couplings in Eq. (\ref{rpvt}). Therefore, R-parity results as an accidental symmetry
originated by the underlying flavor symmetry of the model.
\begin{table}[tb]
\begin{center}
\begin{tabular}{|c|c|c|c|c|c|c|c|}
\hline
 & $\,\hat{L}\,$ & $\,\hat{e}\,$ & $\,\hat{H}_d\,$ & $\,\hat{H}_u\,$ & $\,\hat{Q}\,$ & $\,\hat{d}\,$ & $\,\hat{u}\,$\\
\hline
$\Delta(27)$ & ${\bf 3}$ & ${\bf 1_{1,2,3}}$ & ${\bf 3}^*$ & ${\bf 3}$ & ${\bf 3}$ & ${\bf 1_{1,2,3}}$ & ${\bf 3}$\\
\hline
\end{tabular}\caption{$\Delta(27)$ charges of the MSSM superfields.}
\label{model}
\end{center}
\end{table}
The superpotential invariant under $\Delta(27)$  is given by
\begin{equation} \label{superpot1}
\mathcal{W}_{\text{MSSM}} = Y_u \, \hat{Q} \hat{H}_u \hat{u}\,+Y_d \, \hat{Q} \hat{H}_d \hat{d}\,\,+Y_l \, \hat{L} \hat{H}_d \hat{e}\,+\mu \hat{H}_u\, \hat{H}_d \, ,
\end{equation}
here we have omit the $\Delta(27)$ contractions for simplicity. Note, however, that in the product ${\bf 3}\times {\bf 3}$  there are three
different contractions in the ${\bf 3}^*$ representation. Therefore, each Yukawa
coupling in Eq. \eqref{superpot1} should be understood as three
different parameters, accounting for the three possible $\Delta(27)$
invariant products. We denote them as $Y_{u,d,l}^{1,2,3}$.

Furthermore, the flavor symmetry also forbids the most dangerous
non-renormalizable operators (those with lower dimensions, $d=5$) that
break L or B numbers. These are\footnote{In general, some of these
  operators will not be generated in a given model. For example, the
  operator $\mathcal{O}_7^{(5)}$ will be absent if there is only one
  $\hat H_u$ superfield since the antisymmetric $SU(2)$ contraction
  would vanish exactly.}~\cite{Piriz:1997id}
\begin{align}
\mathcal{O}_1^{(5)} & = [Q Q Q L]_F & \mathcal{O}_2^{(5)} & = [u u d e]_F \nonumber \\
\mathcal{O}_3^{(5)} & = [Q Q Q H_d]_F & \mathcal{O}_4^{(5)} & = [Q u e H_d]_F \nonumber \\
\mathcal{O}_5^{(5)} & = [L L H_u H_u]_F & \mathcal{O}_6^{(5)} & = [L H_d H_u H_u]_F \nonumber \\
\mathcal{O}_7^{(5)} & = [H_u H_u e^*]_D & \mathcal{O}_8^{(5)} & = [H_u^* H_d e]_D \nonumber \\
\mathcal{O}_9^{(5)} & = [Q u L^*]_D & \mathcal{O}_{10}^{(5)} & = [u d^* e]_D \nonumber \\
\mathcal{O}_{11} ^{(5)} & = [Q Q d^*]_D & \nonumber
\end{align}
Here, we include F-terms that may be present in the superpotential and
D-terms which may be present in the K\"ahler potential. They are
denoted with the subscripts $F$ and $D$, respectively. These operators
appear in the lagrangian with a mass suppression $1/\Lambda$, where
$\Lambda$ is the energy scale associated to the L and/or B number
violating physics beyond the MSSM. Even if this scale is taken as
large as the Planck scale, these dangerous operators would lead to too
fast proton decay if the corresponding coefficients are of order
one. Since all these operators are forbidden by $\Delta(27)$, we
conclude that, regarding proton stability, the flavor symmetry makes a
better job than R-parity\footnote{The model in
  Ref. \cite{Carone:1996nd} also leads to R-parity as a
  \emph{by-product} of a flavor symmetry. However, some $d=5$
  operators are allowed by their symmetry. We also note that other
  alternatives to R-parity can also forbid (or strongly suppress)
  higher-dimensional operators with lepton and/or baryon number
  violation, see for example~\cite{Dreiner:2005rd}.}.

One could similarly list all dimension six ($d=6$) non-renormalizable
operators that induce lepton and/or baryon number
violation~\cite{Piriz:1997id}. This list is, of course, much
longer. Although $\Delta(27)$ forbids many of them, some are
allowed. A simple example is $\mathcal{O}^{(6)} = [u d d L H_u]_F$,
which breaks both lepton and baryon numbers.  These operators have a
mass suppression $1/\Lambda^2$ and thus they are less dangerous than
the $d=5$ ones, only requiring relatively small coefficients if
$\Lambda \sim m_{GUT} = 2 \cdot 10^{16}$ GeV~\cite{Piriz:1997id}.

In conclusion, the flavor model given by the $\Delta(27)$ charges in
Table \ref{model} stabilizes the proton (including all dimension 5
dangerous operators) and leads to automatic R-parity
conservation. Furthermore, the $\Delta(27)$ symmetry allows the usual
three fermion Yukawa couplings, as well as the $\mu$ term. Therefore,
we recover the usual MSSM (with the restrictions imposed by the flavor
symmetry) with a clear improvement regarding proton stability. The
next check that we need to make is the viability of the model
regarding fermion masses and mixings.

\section{The quark sector}
\label{sec:quarksector}

In order to get the structure of the fermion mass matrices, we give
here the relevant contractions of the $\Delta(27)$ group. As mentioned
before, the product rules for triplet representations are ${\bf
  3}\times {\bf 3}^*=\sum_{i=1}^9 ({\bf 3}\times {\bf 3}^*)_i \equiv
\sum_{i=1}^9 {\bf 1_i}$. If one denotes the triplet representations as
${\bf 3}=(a_1,~a_2,~a_3)$ and ${\bf 3}^*=(b_1,~b_2,~b_3)$, the
explicit singlet contractions are given by
\begin{eqnarray}
{\bf 1_1}&=&a_1 b_1+a_2 b_2+a_3 b_3, \nonumber \\
{\bf 1_2}&=&a_1 b_1+a_2 b_2\omega+a_3 b_3\omega^2, \nonumber \\
{\bf 1_3}&=&a_1 b_1+a_2 b_2\omega^2+a_3 b_3\omega, \nonumber \\
{\bf 1_4}&=&a_1 b_2+a_2 b_3+a_3 b_1, \nonumber \\
{\bf 1_5}&=&a_1 b_2+a_2 b_3\omega+a_3 b_1\omega^2, \nonumber \\
{\bf 1_6}&=&a_1 b_2+a_2 b_3\omega^2+a_3 b_1\omega, \nonumber \\
{\bf 1_7}&=&a_2 b_1+a_3 b_2+a_1 b_3, \nonumber \\
{\bf 1_8}&=&a_2 b_1+a_3 b_2\omega+a_1 b_3\omega^2, \nonumber \\
{\bf 1_9}&=&a_2 b_1+a_3 b_2\omega^2+a_1 b_3\omega, \nonumber
\end{eqnarray}
with $\omega^3=1$. Similarly, one can obtain the product ${\bf
  3}\times {\bf 3}={\bf 3}^*+{\bf 3}^*+{\bf 3}^*$. If ${\bf
  3}=(a_1,~a_2,~a_3)$ and ${\bf 3}=(b_1,~b_2,~b_3)$, the corresponding
contractions are
\begin{equation}
\left(\begin{array}{c}a_1 b_1\\a_2 b_2\\a_3 b_3\end{array}\right)\oplus
\left(\begin{array}{c}a_2 b_3\\a_3 b_1\\a_1 b_2\end{array}\right)\oplus
\left(\begin{array}{c}a_3 b_2\\a_1 b_3\\a_2 b_1\end{array}\right).
\end{equation} 
The Yukawa term for the $d$ quarks (as well as for the charged
leptons) involves the $\Delta(27)$ product ${\bf 3}\times {\bf 3}^*
\times {\bf 1_i}$ where the singlets are $1_1$, $1_2$ and $1_3$,
respectively. The mass matrix reads
\begin{eqnarray}
M_d &\sim& \left( \begin{array}{ccc}
Y_d^1 \langle H_d^1 \rangle & Y_d^2 \langle H_d^1 \rangle & Y_d^3 \langle H_d^1 \rangle \\
Y_d^1 \langle H_d^2 \rangle & \omega Y_d^2 \langle H_d^2 \rangle & \omega^2 Y_d^3 \langle H_d^2 \rangle \\
Y_d^1 \langle H_d^3 \rangle & \omega^2 Y_d^2 \langle H_d^3 \rangle & \omega Y_d^3 \langle H_d^3 \rangle
\end{array} \right) \label{mass-ld} 
\end{eqnarray}
The $u$ quark mass matrix comes from the $\Delta(27)$ product ${\bf
  3}\times {\bf 3}\times{\bf 3}$, which leads to
\begin{eqnarray}
M_u &\sim&  \left( \begin{array}{ccc}
Y_u^1 \langle H_u^1 \rangle & Y_u^2 \langle H_u^2 \rangle & Y_u^3 \langle H_u^1 \rangle \\
Y_u^3 \langle H_u^2 \rangle & Y_u^1 \langle H_u^2 \rangle & Y_u^2 \langle H_u^3 \rangle \\
Y_u^2 \langle H_u^1 \rangle & Y_u^3 \langle H_u^3 \rangle & Y_u^1 \langle H_u^3 \rangle
\end{array} \right) \label{mass-u}
\end{eqnarray}
It is not difficult to show that by taking~\footnote{The alignment
  $\langle H_{u,d}^0 \rangle \sim (1,1,1)$ will be assumed as a
  starting point in the derivation of the fermion mixing
  matrices. Although we do not address its origin, we note that this
  alignment is natural in models based on the $\Delta(27)$ discrete
  symmetry~\cite{deMedeirosVarzielas:2006fc}. Nevertheless, small
  deviations from this particular alignment do not change our main
  conclusions and would be welcome in order to obtain a realistic CKM
  matrix (see below).} the vaccum expectation value (VEV) alignment
$\langle H_{u,d}^0 \rangle \sim (1,1,1)$, which breaks $\Delta(27)$
into a $Z_3$ subgroup,
all the charged fermion mass matrices are diagonalized on the left by
the same unitary matrix, the so-called {\it magic} matrix $U_\omega$,
defined as
\begin{equation}\label{Uom}
U_\omega =\frac{1}{\sqrt{3}} 
\left( 
\begin{array}{ccc}
1& 1&1\\
1&\omega&\omega^2\\
1&\omega^2&\omega\\
\end{array} 
\right) \, ,
\end{equation}
namely
\begin{equation}\label{Uom}
U_\omega^\dagger \, M_u\cdot M_u^\dagger \,U_\omega = D^2_u\,,\qquad
U_\omega^\dagger \, M_d\cdot M_d^\dagger \,U_\omega = D^2_d
\end{equation}
where $D_{u,d}$ are diagonal matrices whose entries are functions of
the Yukawa couplings for each sector. Then, each charged fermion mass
matrix in Eqs. \eqref{mass-ld} and \eqref{mass-u} can be fitted to
reproduce the corresponding three fermion masses and the CKM mixing
matrix, $V_{CKM}=U_{u}^\dagger U_{d}$, turns out to be proportional to
the identity, since $U_{u}=U_{d}=U_\omega$. This is usually regarded
as a good starting point when building a model. As next step one can
consider a completely broken $\Delta(27)$, breaking the alignment
$\langle H_{u,d}^0 \rangle \sim (1,1,1)$, and leading to a CKM that
deviates from the identity. Similarly, the CKM matrix can also be
generated at the loop level from the flavor structure associated with
the SUSY breaking terms~\cite{Babu:1998tm}. Other possibilities to
generate the CKM mixing could be by means of adding extra mirror
quarks, scalar-mediated interactions \cite{Lavoura:2007dw} or by using
different singlets of $\Delta(27)$ \cite{Bhattacharyya:2012pi}.

We now turn to the next section, where we discuss extensions of the
lepton sector to generate non-zero neutrino masses. Although one may
expect to find many possible directions, we will find that this task
turns out to be far from trivial due to the restrictions imposed by
$\Delta(27)$.

\section{The lepton sector}
\label{sec:leptonsector}

\subsection{Charged lepton}
\label{subsec:charlepton}

The charged lepton mass matrix has the same structure as the $d$ quark mass matrix, that is 
\begin{eqnarray}
M_{l} &\sim& \left( \begin{array}{ccc}
Y_{l}^1 \langle H_d^1 \rangle & Y_{l}^2 \langle H_d^1 \rangle & Y_{l}^3 \langle H_d^1 \rangle \\
Y_{l}^1 \langle H_d^2 \rangle & \omega Y_{l}^2 \langle H_d^2 \rangle & \omega^2 Y_{l}^3 \langle H_d^2 \rangle \\
Y_{l}^1 \langle H_d^3 \rangle & \omega^2 Y_{l}^2 \langle H_d^3 \rangle & \omega Y_{l}^3 \langle H_d^3 \rangle
\end{array} \right), \label{mass-l} 
\end{eqnarray}
therefore if the vaccum expectation value (VEV) alignment $\langle
H_{u,d}^0 \rangle \sim (1,1,1)$, this matrix is also diagonalized by
the $U_{\omega}$ matrix given in Eq. \eqref{Uom}.

\subsection{Neutrino masses -- inverse type-III seesaw}
\label{subsec:numass}

So far, we have only discussed the viability of the
framework. Unfortunately, the model resulting from the addition of the
$\Delta(27)$ discrete symmetry is just the MSSM, with some
restrictions in the parameters\footnote{To be precise, the only
  difference with the canonical MSSM is the existence of three pairs
  of Higgs doublets, as discussed below.}. Therefore, we do not expect
any new collider signature that is not present in the canonical
MSSM. In order to find new predictions one needs to extend the model
in order to account for neutrino masses. The new structures must
preserve $\Delta(27)$ as well, and this leads to interesting
consequences.

As shown above, the flavor symmetry forbids the Weinberg operator,
$\mathcal{O}_5^{(5)} = [L L H_u H_u]_F$.  Therefore, in order to
generate neutrino masses one is forced to go beyond minimal models and
consider higher dimensional operators.  

To the particle content in Table \ref{model}, we add the superfields in
Table\,\ref{model2}. The singlet $\hat S$ should not be confused with
the NMSSM singlet superfield.  In fact, note that the $\hat{S}
\hat{H}_u \hat{H}_d$ superpotential term is forbidden by $\Delta(27)$.
Besides $\mathcal{W}_{MSSM}$ defined in Eq.\,\eqref{superpot1} the
superpotential contains
\begin{eqnarray}
\mathcal{W} \supset
Y_\Sigma \hat{L} \hat{H}_u \hat{\Sigma}_1 + M \hat{\Sigma}_1 \hat{\Sigma}_2 + \lambda \hat{S} \hat{\Sigma}_2 \hat{\Sigma}_2 
+\kappa_S \hat{S}^3
\end{eqnarray}
Other superpotential terms are forbidden by the
gauge and flavor symmetries\footnote{We note that if the
  $\hat{\Sigma}_{1,2}$ were singlets under $SU(2)_L$, superpotential
  terms $\kappa_{\Sigma_i} \hat{\Sigma}_{i}^3$ would be allowed, thus
  breaking R-parity explicitly.}.
\begin{table}[tb]
\begin{center}
\begin{tabular}{|c|c|c|c|}
\hline
 & $\,\hat{\Sigma}_1\,$ & $\,\hat{\Sigma}_2\,$ &$\,\hat{S}\,$\\
\hline
$\Delta(27)$ & ${\bf 3}$ & ${\bf 3}^*$ & ${\bf 3}^*$\\
\hline
\end{tabular}
\caption{Charge assignment of the additional superfields in the
  extended model for neutrino masses. The superfields
  $\hat{\Sigma}_{1,2}$ are triplets of $SU(2)_L$ with $Y=0$ and
  $\hat{S}$ is a neutral $SU(2)_L$ singlet.}
\label{model2}
\end{center}
\end{table}
When the scalar component of $\hat S$ gets a VEV, an effective
Majorana mass for the $\Sigma_2$ triplet (the fermionic component of
the $\hat{\Sigma}_2$ superfield) is generated.  This leads to an
inverse seesaw mechanism~\cite{Mohapatra:1986bd} induced by $SU(2)_L$
triplets (for other realizations of the \emph{inverse type-III seesaw}
see~\cite{Ma:2009kh}). In the basis $\psi^T =
(\nu,\,\Sigma_1^0,\,\Sigma_2^0)$, we obtain the $9\times 9$ mass
matrix for the neutral fermions
\begin{equation}
M_\nu=
\left(
\begin{array}{ccc}
0 & Y_\Sigma \vev{H_u}& 0 \\
Y_\Sigma^T \vev{H_u} & 0 & M\\
0&M^T&\lambda \vev{S}
\end{array}
\right) \, ,\label{seesaw1}
\end{equation}
which, assuming $\lambda v_S \ll Y_\Sigma v_u \ll M$, leads to
\begin{equation}
m_\nu= v_u^2 v_S Y_\Sigma (M^T)^{-1} \lambda M^{-1} Y_\Sigma^T 
\label{seesaw2}\end{equation}
where $v_u=\vev{H_u^0}$ and $v_S = \vev{S}$.
In fact, it is worth emphasizing some advantages that our model has
with respect to the conventional inverse seesaw. Typically, in the
context of inverse seesaw models it is quite difficult to forbid the
$\Sigma_1^0 \Sigma_1^0$ mass term after ones allows for lepton number
violation. Here its absence is a direct consequence of the
$\Delta(27)$ symmetry. Moreover, in order to have new physics at the
TeV scale, that is $M\sim$TeV, and $\mathcal{O}(1)$ Yukawa couplings,
the parameter $\mu_S = \lambda v_S$ must be of the order of $10$ eV. In
the original inverse seesaw mechanism the $\mu_S$ parameter is expected
to be naturally small in the 't Hooft sence since the limit $\mu_S \to
0$ enhances the symmetry of the lagrangian (lepton number is
recovered). Although this is not the case here, $\mu_S$ can be
suppressed by means of the VEV of the scalar $S$. Moreover, several
mechanisms to obtain dynamically a small $\mu_S$ parameter are known in
the literature~\cite{Ma:2009gu}. For these reasons, we consider a
small $\mu_S$ parameter a natural choice.

\begin{figure}[tb]
\centering
\includegraphics[width=0.6\linewidth]{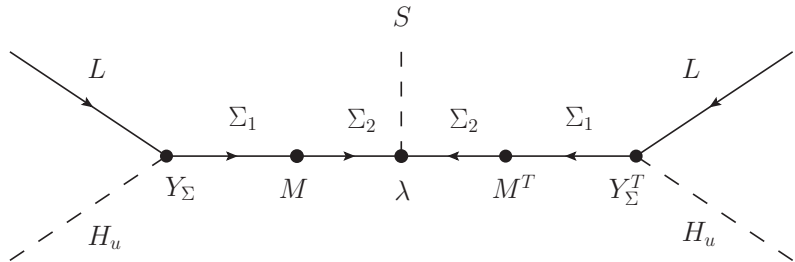}
\caption{$\mathcal{O}^{(6)} = [L L H_u H_u S]_F$ realization.}
\label{mnu-dim6}
\end{figure}

As mentioned above, the flavor symmetry forbids the Weinberg
operator. In fact, in this model neutrino masses are generated due to
the effective operator
\begin{equation} \label{op6}
\mathcal{O}^{(6)} = [L L H_u H_u S]_F \, ,
\end{equation}
which is obtained after integrating out the $\hat{\Sigma}_{1,2}$
superfields, as depicted in Figure\,(\ref{mnu-dim6}). Therefore, the
proposed model can be seen as a concrete renormalizable realization of
this operator, which is the minimal implementation of Majorana
neutrino masses after the Weinberg operator is forbidden.

Lepton number is explicitly broken by the trilinear superpotential
term $\kappa_S \hat{S}^3$ and thus no majoron, the Goldstone boson
associated to the spontaneous breaking of $U(1)_L$, appears in the
spectrum. However, R-parity remains unbroken after the addition of
this term, since $S$ carries two units of lepton number.
One may also wonder about the appearance of new unwanted interactions
after the introduction of the $\hat S$ superfield. It is
straightforward to check that no L or B violating renormalizable
operators are allowed, apart from the aforementioned $\kappa_S
\hat{S}^3$ term. Regarding higher dimensional operators, we note that
the operator $\mathcal{O}^{(5)} = [u d d S]_F$ is the only dimension 5
operator (with L or B violation) that is allowed by the
symmetries. However, it cannot be generated at tree-level, since $\hat
S$ does not couple to the quark superfields, and only some specific
high-energy completions of our model would generate it at the (less
dangerous) loop level.

\subsection{Neutrino mixing pattern}
\label{sec:numix}

Let us now discuss the resulting neutrino mixing pattern. Using the
$\Delta(27)$ contraction rules, one finds that the mass matrix $M$ is
proportional to the identity matrix. The Dirac neutrino mass matrix
$m_D=Y_\Sigma \vev{H_u^0}$ and the matrix $\mu_S = \lambda \vev{S}$ have
the structures
\begin{eqnarray}
m_D=
\left(
\begin{array}{ccc}
\alpha & \beta & \gamma \\
\gamma & \alpha & \beta\\
\beta&\gamma&\alpha
\end{array}
\right)\,,\,
\mu_S=\left(
\begin{array}{ccc}
\alpha^\prime& \beta^\prime & \beta^\prime \\
\beta^\prime & \alpha^\prime &\beta^\prime\\
\beta^\prime&\beta^\prime&\alpha^\prime
\end{array}
\right)
\end{eqnarray}
when all the scalar fields take VEV in the $(1,1,1)$ direction. In
this case, the charged lepton mass matrix\footnote{After electroweak
  symmetry breaking, charged leptons mix with the charged components
  of the $\Sigma$ fermions. This leads, however, to numerically
  irrelevant corrections in the leptonic mixing matrix.} is
diagonalized by the magic matrix $U_\omega$, see Eq.~\eqref{Uom}. One
can now perform a $U_\omega$ rotation in order to go to the basis
where the charged lepton mass matrix is diagonal. In this basis, the
neutrino mass matrix can be written as
\begin{equation}\label{mass2mass}
\tilde{m}_\nu\sim U_{\omega}\,m_\nu\, U_{\omega}^T\sim \left(
\begin{array}{ccc}
 a & 0 & 0 \\
 0 & 0 & b \\
 0 & b & 0
\end{array}
\right)
\end{equation}
where the parameters $a$ and $b$ are functions of
$\alpha,\,\beta,\,\gamma,\alpha',\,\beta',\,\gamma'$.  In general $a
\ne b$. By setting $a=b$ we recover the result of the Babu-Ma-Valle
model~\cite{Babu:2002dz}. In this limit, the mass matrix in
Eq.~\eqref{mass2mass} gives maximal atmospheric angle and degenerate
neutrino mass spectrum, while the solar and reactor mixing angles are
zero. As observed in \cite{Babu:2002dz}, Eq.~\eqref{mass2mass} is
corrected by wave-function renormalizations of $\nu_e$, $\nu_\mu$, and
$\nu_\tau$, as well as the corresponding vertex renormalizations,
lifting the neutrino degeneracy and the solar/reactor mixing
angles. The resulting neutrino mass matrix can be written as
\begin{small}
\begin{equation}\nonumber
\tilde{m}_\nu^{\text{1-loop}}\sim \left( 
\begin{array} {c@{\quad}c@{\quad}c} 
a (1 + 2 \delta_{ee}) & a \delta_{e \mu} + b \delta_{e \tau}^* & b \delta_{e \mu}^* + a \delta_{e \tau} \\
 & 2 b \delta_{\mu \tau}^* & b (1 + \delta_{\mu \mu}+\delta_{\tau \tau}) \\ 
 & & 2 b \delta_{\mu \tau} \end{array} \right)\,,
\end{equation}
\end{small}
where $\delta_{ij}$ parametrize the radiative corrections. When $a=b$
(as in \cite{Babu:2002dz}) and assuming real $\delta_{ij}$
corrections, the resulting neutrino mass matrix is $\mu-\tau$
symmetric giving maximal atmospherc angle and zero $\theta_{13}$
reactor angle. The solar angle is a free parameter and can be fitted.
If the corrections are allowed to be complex, a non-zero value for
$\theta_{13}$ can be obtained. In this case, CP violation is predicted
to be maximal~\cite{Babu:2002dz}. Regarding the nature of the
corrections, these come from flavor mixing in the slepton/sneutrino
sector. A detailed study can be found
in~\cite{Hirsch:2003dr}.  Note that our model has more freedom
since $a \ne b$ in general. This can be used to relax some of the
restrictions in the parameter space. Nevertheless, large
$\tilde{\mu}-\tilde{\tau}$ mixing is necessary, typically predicting
Br($\tau \to \mu \gamma$) close to its experimental
limit. Furthermore, in the SUSY inverse seesaw one expects large
Z-penguin contributions in lepton flavor violating
processes~\cite{Hirsch:2012ax}.  Therefore, observables such as
Br($\tau \to \mu \ell \ell$), with $\ell = e, \mu$, are also expected
to set important constraits on the SUSY parameter space. Finally,
violation of lepton flavor universality in observables such as $R_K$
and $R_\pi$ is also an important test of the
model~\cite{Abada:2012mc}.

\section{Summary and discussion}
\label{sec:summary}

In summary, we have proposed a supersymmetric flavor model based on
the $\Delta(27)$ symmetry. All lepton and/or baryon number violating
operators of dimension 4 and 5 are forbidden by the flavor symmetry,
thus providing a single explanation for the proton and dark matter
stabilities\footnote{For dark matter stability from the spontaneous
  breaking of a non-Abelian discrete flavor symmetry
  see~\cite{Hirsch:2010ru}.}. The extension to account for neutrino
masses and mixings turned out to be non-trivial due to the
restrictions imposed by the $\Delta(27)$ symmetry. In fact, naive
extensions of the leptonic sector typically spoil the nice features of
the original model and depart from the main motivations for this
work. We found that one can indeed generate neutrino masses by means
of an inverse type-III seesaw, while preserving R-parity and keeping
the proton stable. The phenomenology of the complete model has been
briefly discussed. In particular, the observed neutrino mixing pattern
(with a large reactor angle) can be well accommodated. 

Before concluding the paper we would like to comment on a
difference that our setup has with respect to the canonical MSSM: the
existence of three Higgs doublets. This feature, the replication of
the minimal Higgs sector, is shared by all flavor models where the
usual Higgs doublet is promoted to a multiplet of the flavor group.
This typically implies very complicated scalar potentials, whose
minimization leads to viable vacua (from a phenomenological point of
view) thanks to the soft SUSY breaking terms. An example in the
context of the $A_4$ symmetry can be found in
Ref. \cite{Bazzocchi:2012ve}. Although the model under consideration
here involves a different flavor symmetry, similar results are
expected. A detailed minimization of the potential and determination
of the scalar spectrum is clearly beyond the scope of the
paper. Furthermore, if the new scalar/pseudoscalar states can be made
relatively light, one may expect sizable Higgs-mediated contributions
to flavor processes such as $\tau \to 3 \mu$, $B_{d,s} \to \ell_i
\ell_j$ or $\tau \to \mu P$ (where $P$ denotes a neutral pseudoscalar
meson)\footnote{The magnitude of these contributions will depend, of
  course, on the value of $\tan \beta$.}~\cite{Dedes:2002rh}.

The extended Higgs sector provides additional freedom to accommodate
the observed $126$ GeV Higgs-like resonance found by ATLAS and CMS. In
fact, one can easily find regions in parameter space very similar to
the so-called \emph{decoupling limit} of the MSSM, where the lightest
scalar particle behaves as the SM Higgs boson. We note that the
non-standard corrections coming from the triplets are typically very
small due to the relatively small Yukawa couplings (see for example
Ref. \cite{Elsayed:2011de}).

Although most of the phenomenological features of our setup can be
found in other flavor models, their combination is quite
distinctive. In particular, the presence of the light $\Sigma_{1,2}$
$SU(2)_L$ fermion triplet leads to a very rich collider phenomenology,
since it can be pair produced at the LHC due to its gauge
interactions. This possibility, not present in singlet extensions of
the MSSM, may lead to very clear signatures with additional leptons
and/or lepton flavor violation in the $\Sigma_{1,2}$ decays
\cite{Franceschini:2008pz}.  It is well-known that the assumption of
R-parity in the MSSM serves to stabilize the LSP. Here we obtain the
same result, \emph{R-parity}, without assuming it. Therefore, the
usual MSSM candidate, a neutralino LSP, can play the role of the dark
matter of the universe.

\section*{Acknowledgements}

The authors are grateful to Werner Porod for useful comments on the
manuscript.
This work was supported by grants FPA2011-22975, MULTIDARK
CSD2009-00064, Prometeo/2009/091, EU ITN UNILHC PITN-GA-2009-237920,
S.M has been also supported by DFG grant WI 2639/4-1.
A.V. acknowledges support by the ANR project CPV-LFV-LHC {NT09-508531}. 

\end{document}